\documentstyle [twocolumn,prl,aps,floats,graphicx] {revtex}

\begin{document}
\draft
\flushbottom
\begin{title}
{\bf Thermal quenching of electronic shells and channel competition 
in cluster fission}
\end{title} 
\author{Constantine Yannouleas and Uzi Landman} 
\address{
School of Physics, Georgia Institute of Technology,
Atlanta, Georgia 30332-0430, USA}
\author{C. Br\'{e}chignac, Ph. Cahuzac, B. Concina, and
J. Leygnier}
\address{Laboratoire Aim\'{e} Cotton, CNRS B\^{a}t. 505, Campus Orsay,
91405 Orsay cedex, France}

\date{Phys. Rev. Lett. {\bf 89}, 173403 (2002)}
\maketitle

\begin{abstract}
Experimental and theoretical studies of fission of doubly-charged 
Li, Na, and K clusters in the low fissility regime reveal the strong 
influence of electronic shell effects 
on the fission products. The electronic entropy controls the 
quenching of the shell effects and the competition between 
magic-fragment channels, leading to a transition from favored channels of 
higher mass symmetry to the asymmetric channel involving the trimer cation at 
elevated temperatures.
\end{abstract}

\narrowtext

Explorations of factors controlling the stability and the collapse or 
fragmentation of finite systems range from early studies of 
hydrodynamical instabilities and droplet formation \cite{ray} to studies
of fission and fragmentation of atomic nuclei \cite{bm}, and of atomic and
molecular free clusters and supported nanostructures \cite{lan,bre1}. 
Here we focus on the charge instability and fission of doubly charged 
cationic metal clusters, that is
${\text{M}}_N^{2+} \rightarrow {\text{M}}_P^+ + {\text{M}}_{N-P}^+$. Such 
instabilities in macroscopic fluid systems have been studied by Rayleigh
\cite{ray}, and adaptations of Rayleigh's model (RM) for the description and 
analysis of nuclear \cite{bm} and cluster fission \cite{sau} are referred to 
as the Liquid Drop Model (LDM). In the LDM, the 
clusters are viewed as charged classical liquid drops whose shape and 
dynamics are controlled by the competition between the repulsive Coulomb 
$(E_{\text{Coulomb}})$ and the binding surface ($E_{\text{surface}}$) 
energies \cite{yl1}. 

Various fission regimes may be classified by the fissility parameter
$X = E^{\text{sph}}_{\text{Coulomb}}/2 E^{\text{sph}}_{\text{surface}}$,
where the superscript ``sph'' indicates that the
equilibrium shapes of the LDM are spherical. For $X \geq 1$,
spontaneous barrierless fission is predicted and observed \cite{bre2,cha};
this is also the regime studied originally by Rayleigh who predicted
a mass-symmetric fission mode (see also Ref.\ \cite{bre2}). In the $X < 1$ 
regime, which is the focus of our paper, the fissioning system must overcome a 
barrier. For atomic clusters in this regime, tunneling is suppressed,
unlike the case of nuclei, and fission requires thermal activation. From 
$X_{\text{cr}}=1$, the RM critical size for barrierless fission is 
$N^{\text{RM}}_{\text{cr}}=z^2e^2/(16\pi r_s^3 \sigma)$, where $r_s$ is the
Wigner-Seitz radius and $\sigma$ is the surface-energy; for doubly
charged $(z=2)$ Li, Na, and K clusters, one has $N^{\text{RM}}_{\text{cr}} 
\approx 10$. Thus, with the exception of the smallest ones, fissioning 
clusters must be hot \cite{bre3} in order to overcome the 
fission barrier. 

For the low fissility $X < 1$ regime, the LDM predicts the dominance of a 
strongly mass asymmetric fission process, as deduced from the 
finite-temperature \cite{yl1,yl2} $Q_P^T=F_{P}^+ + F_{N-P}^+ - F_{N}^{2+}$
values, that express the free-energy balance for the possible $(P, N-P)$ 
fission channels. Such asymmetric fission has been observed in experiments for
Na$_N^{2+}$  with $N \sim 24$ (i.e., near the appearance size, see below), 
where the preferred fission channel involved the ``magic'' 
Na$_3^+$ fragment \cite{bre3}. The prevalence of this channel, 
compared to other asymmetric ones in the neighborhood of the LDM minimum, 
originates from quantum mechanical shell effects \cite{raj}.

In this paper we present experimental and theoretical results,
pertaining to cluster fission for $X << 1$, which cannot be described by the 
classical RM, demonstrating instead the dominance of quantum-size effects, 
i.e., electronic shell contributions \cite{yl1}. In particular, we study 
alkali-metal (Li, Na, and K) clusters with $N=24$, 26, 28, and 30. These 
systems correspond to the lowest observable fissility regime with 
$N >> N_{\text{cr}}^{\text{RM}}$; e.g., for $N=30$ one has 
$X \approx 0.25$ for the doubly charged alkali clusters. Moreover, this range 
of sizes is particularly suitable for explorations of shell effects, since it 
includes cases where the clusters may fission into doubly ``magic'' fragments,
i.e., ${\text{M}}_{24}^{2+} \rightarrow {\text{M}}_{21}^{+} + 
{\text{M}}_{3}^{+}~\;{\text{and}}\;~ 
{\text{M}}_{30}^{2+} \rightarrow {\text{M}}_{21}^{+} + {\text{M}}_{9}^{+}$.

We find that, in addition to M$_3^+$, the more mass symmetric magic fission 
channels (involving the magic fragments M$^+_9$ and M$^+_{21}$) do appear and 
compete with the channel involving the trimer (even becoming the preferred 
channels in several instances). Furthermore,
from comparisons between the experiments and calculations based on the
finite-temperature shell-correction method (FT-SCM) \cite{note4}, 
we find that the channel distributions are greatly influenced by thermal 
quenching of the shell effects due to the increase (compared to $T=0$) of the 
electronic entropy; this quenching results from thermal promotion of 
electrons to unoccupied single-particle levels which smears out the shell 
structure. The more mass-symmetric fission channels are
favored at lower temperatures, whereas the asymmetric ones (e.g., involving
the M$_3^+$ fragment) dominate as $T$ increases. These trends reflect the
more pronounced (quenching) effect of the electronic entropy on the
shell-stabilization of the larger magic sizes (e.g., M$_9^+$ and M$_{21}^+$).

In the experiment \cite{bre4},
a distribution of relatively large neutral clusters is
formed by a gas aggregation source. The clusters are ionized, photo-excited 
and warmed by a 15 ns laser pulse (Nd-YAG laser at $h \nu =3.50$ eV). 
A rapid sequential evaporation follows the warming of the clusters for about 
1 $\mu$s (i.e., the residence time in the ionizing region). 
This results in a charged cluster distribution shifted down to smaller sizes, 
called an ``evaporative ensemble'' (EE) \cite{bre4}. The cluster temperature 
associated with the EE is determined \cite{bre4} by the 
cluster dissociation energy and by the experimental time window, i.e., the 
residence time in the ionizing/heating region. 
Under our experimental conditions, the temperatures of the EE's of lithium, 
sodium and potassium clusters are 700$\pm$100 K, 400$\pm$100 K and 
300$\pm$100 K, respectively. 

\begin{figure}[t]
\centering\includegraphics[width=8.0cm]{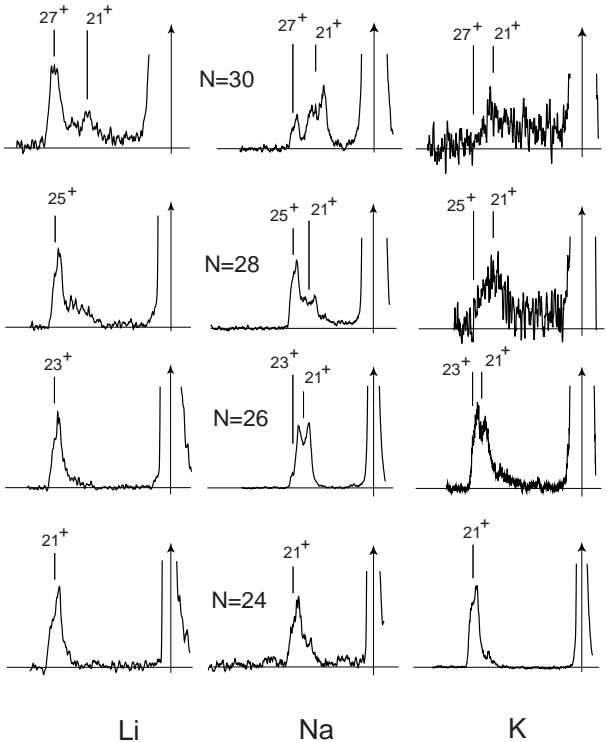}\\
~~~\\
\caption{
Fission channels recorded in TOF measurements for Li$_N^{2+}$, Na$_N^{2+}$, 
and K$_N^{2+}$ clusters, with $N=24$, 26, 28, and 30. In each case, the peak
on the right (marked by an arrow) corresponds to the parent cluster and the 
ones on the left of it correspond to the larger fission fragments. Fragment 
labelings are determined from the voltage setup. Peak heights are in arbitrary
units. The splitting of the TOF ion-fragment peaks are due to 
the kinetic energy released during fission, which is of the order of 1 eV 
(see, e.g., the case $N = 26$).
}
\end{figure}

The EE of ionized clusters enters a tandem time-of-flight 
(TOF) mass spectrometer. Parents of interest are mass selected in the first 
TOF region. Because they are at a given temperature, subsequent 
unimolecular dissociation of these parents takes place and the resulting 
fragments are mass analyzed by the second TOF device. Enhanced optimization of
the electrostatic potentials in the current experiments has led to significant
increases in both the mass and energy resolutions. In particular, the
latter allows us to resolve the kinetic energy released during the fission 
process (see the splitting of the peak of some of the fission fragments 
in Fig.\ 1).

For cluster sizes in the vicinity of the appearence size (see below), 
the unimolecular dissociation of doubly charged clusters portrays 
the competition between the fission (1a) and evaporation (1b) processes:
\begin{mathletters}
\begin{eqnarray}
& {\text{M}}_N^{2+} & \rightarrow  {\text{M}}_{P}^+   +  {\text{M}}_{N-P}^+
 \\
& ~~~ & \rightarrow  {\text{M}}_{P'}^{2+}   +  {\text{M}}_{N-P'}~,
\; P'=1 \; {\text{or}} \;2~.
\label{dis}
\end{eqnarray}
\end{mathletters}
We focus here on the fission decay channels. The explored size domain is 
bounded by two values, $N_-$ and $N_+$. The lower value $N_-$ is determined 
by the appearance size $N^{\text{app}}$ ($N^{\text{app}}=24$ for Li$_N^{2+}$ 
and Na$_N^{2+}$ clusters, and 19 for K$_N^{2+}$ clusters \cite{bre1}), below 
which fission is the dominant dissociation process [see Eq.\ (1a)]. Below 
this size no doubly charged clusters are present in the mass spectra,
since their fast fission process prevents their observation during the 
experimental time window. $N_+$ is determined by the loss of the signal 
corresponding to the fission process, resulting from the increase of the
inner part of the fission barrier with the size of the cluster relative to the
essentially constant monomer evaporation energy.
Fig.\ 1 summarizes the results obtained for the three alkali metals in the 
same size range. 

Although we have recorded the fragmentation spectra for all the sizes 
$24 \leq N \leq 30$, only results for clusters with an even number of atoms 
are shown (due to pairing they exhibit simpler features). In our 
experimental setup, only the heavy fragments M$_{P}^+$ with $P > N/2$ are 
detected. The main features are as follows: For the lowest parent size 
($N=24$) and for all three elements, only one fission channel is present,
i.e., the doubly magic $M_3^+ + M_{21}^+$. Increasing the parent size,
however, results in differentiation between the three elements regarding the 
observed fission channels. A larger parent size (i.e., $N \geq 26$) opens the 
channels of heavier fragments and this trend is more pronounced for potassium 
than for lithium. In particular, the trimer appears to be the dominant product
in the Li fission, while M$_{21}^+$ is the dominant one for K, with Na 
exhibiting an intermediate behavior. That is, the asymmetry of the fission 
process is reduced for the K clusters as compared to the Na and Li ones.  

\begin{figure}[t]
\centering\includegraphics[width=8.0cm]{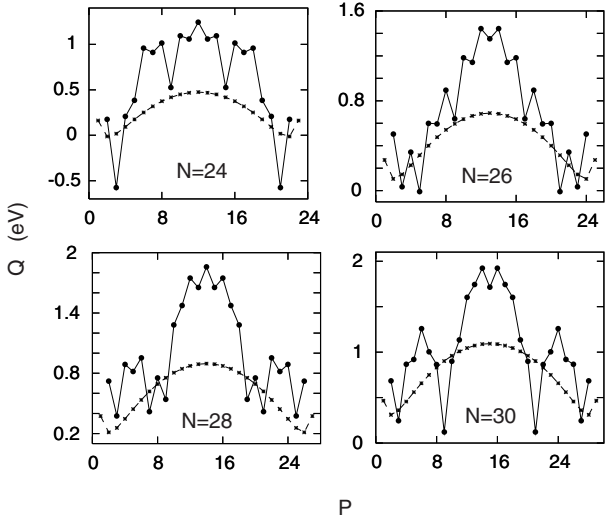}\\
~~~\\
\caption{
Theoretical SCM zero-temperature energetics ($Q$-values) for the processes
Li$_{N}^{2+}$ $\rightarrow$ Li$_P^+$ $+$ Li$_{N-P}^+$, 
$N=24$, 26, 28, and 30. The dashed lines correspond to LDM calculations.
In each case, the horizontal axis corresponds to various SCM channels
$2 \leq P \leq N-2$. 
}
\end{figure}

We have performed FT-SCM 
calculations of the $Q^T$-values for the various fission channels for all 
three alkali-metal sequences at the corresponding experimental 
temperatures. Since for each species the Coulomb repulsion (QR) at the scission
point (measured by the energy release) is approximately constant for all
fission channels in the size-range considered here $(X << 1)$, the effective
inner part of the fission barrier can be approximated by adding the calculated
$Q^T$-values to the QR term. Thus the ordering of the fission channels can be 
directly inferred from the $Q^T$-values. 

The theoretical zero-temperature $Q$-values provide an inadequate description 
of the experimental data. For example, the $Q$-values for Li$_{N}^{2+}$ 
(plotted in Fig.\ 2) cannot explain the observed trend (see Fig.\ 1, left 
column) of strong asymmetric fission, with the Li$_3^+$ being the preferred 
channel for all parents $N=24$, 26, 28, and 30 \cite{note2}. In particular, 
the $T=0$ theoretically deduced preferred fission channels for $N=26$ and 
$N=30$ are (5,21) and (9,21), respectively, in contradiction with the 
measurements (see Fig.\ 1). In contrast, the theoretical $Q^T$-values   
(Fig.\ 3) agree well with the experiment. Specifically, for Lithium the 
calculations at $T$ $=$ 700 K show that Li$_3^+$ is indeed the preferred 
channel for all parents; only for Li$_{30}^{2+}$, the Li$_{21}^+$ channel 
becomes competitive, remaining however less abundant than the one involving 
the trimer.

\begin{figure}[t]
\centering\includegraphics[width=7.5cm]{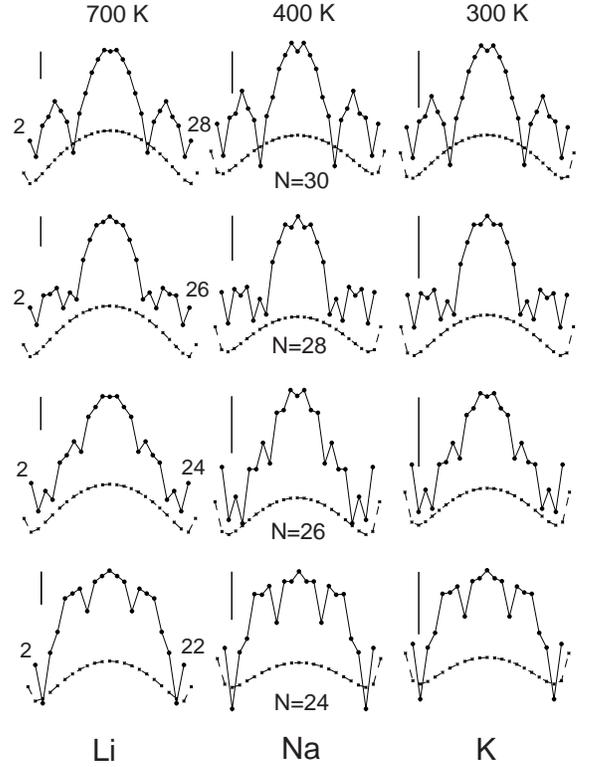}\\
~~~\\
\caption{
Finite-temperature SCM results ($Q^T$-values) for the fission process
M$_N^{2+}$ $\rightarrow$ M$_P^+$ $+$ M$_{N-P}^+$ for Li$_N^{2+}$, 
Na$_N^{2+}$, and K$_N^{2+}$ with $N=24$, 26, 28, and 30.
The corresponding temperatures (in degrees Kelvin), taken from the
experiments, for each species are shown at the top. 
The dashed lines correspond to LDM calculations.
In each case, the horizontal axis corresponds to various SCM channels
$2 \leq P \leq N-2$. The length of the vertical bar in each frame 
corresponds to 0.4 eV.
}
\end{figure}
We reiterate that in Fig.\ 3 there is a clear tendency toward opening the 
channels involving the heavier magic fragments ($P=9$ and $P=21$), both when 
going up vertically from the lighter to the heavier parentss (from $N=24$ to 
$N=30$), as well as when going horizontally from left to right (i.e., from Li 
to K, with Na representing an intermediate case).

\begin{figure}[t]
\centering\includegraphics[width=8.0cm]{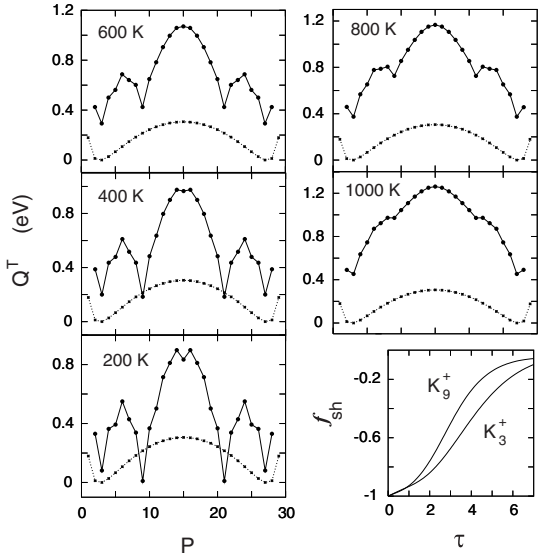}\\
~~~\\
\caption{
Theoretical FT-SCM temperature dependence of the $Q^T$-values for the
process K$_{30}^{2+}$ $\rightarrow$ K$_P^+$ $+$ K$_{30-P}^+$.
The dashed lines correspond to LDM calculations.
In each case, the horizontal axis corresponds to various SCM channels
$2 \leq P \leq 28$. In the bottom right frame, we display $f_{\text{sh}}$
vs. $\tau$ (see text) for the K$_3^+$ and K$_9^+$ fragments.
}
\end{figure}

The physics underlying this agreement between the experiment and the FT-SCM 
results can be revealed by an examination of the influence of the electronic 
entropy on the various fission channels $(P,N-P)$ as a function of $P$. For 
brevity, we limit our discussion to the case of the K$_{30}^{2+}$ parent; the 
conclusions, however, extend to all three species and the other parent sizes. 
Figure 4 displays the FT-SCM $Q^T$-values for K$_{30}^{2+}$ at five 
temperatures ($200$ K to $1000$ K). 
At the lowest temperature, the $(9,21)$ channel is the preferred one; however,
an increase in $T$ is accompanied by 
quenching of the shell effects; i.e., at high temperature (1000 K) the 
shell effects have almost vanished, and the $Q^T$ curve exhibits a 
smooth shape similar to that of the LDM (although with a steeper variation) 

Most important for the competition between the various magic fragments, 
however, is the fact that the electronic-entropy quenching of the shell
effects is non-uniform, with the higher occupied shells of the heavier 
fragments being influenced the most as $T$ increases, since the energy
spacing between the major electronic shells,
$\hbar \omega_0 (N) = 49\; {\text{eV}} \cdot a_0^2 /(r_s^2 N^{1/3})$,
decreases with larger $N$. Consequently, a higher $T$ favors the trimer 
(M$_3^+$) channel over those involving the magic M$_9^+$ and M$_{21}^+$ 
fragments. This happens even in the case of $N=30$ where the competing 
$(9,21)$ channel is doubly magic. Note that, unlike the $T=700$ K experimental
temperature of Li, the temperatures of 300 K for Potassium and 400 K for 
Sodium are low enough so that the heavier-fragment channels retain their $T=0$
original prevalence for all three parent sizes $N=26$, 28, and 30.

Further insight can be gained by examining 
the ratio $f_{\text{sh}}(N) \equiv \Delta F_{\text{sh}}(N,\tau) / 
|\Delta F_{\text{sh}}(N,0)|$ of the free-energy shell-correction term at 
$\tau = 2 \pi^2 k_B T/\hbar \omega_0(N)$ over the shell correction term at 
$\tau=0$. For a schematic single-particle spectrum consisting of equally 
spaced major shells with the same degeneracy, it can be shown in the case of 
closed shells (see p. 608 of Ref.\ \cite{bm} and Ref.\ \cite{yl2})  
that $f_{\text{sh}}(N) \approx -\tau \exp(-\tau)$ for $\tau \geq 1$. Thus,
because of the large prefactor $2\pi^2$, the electronic-entropy effects
manifest themselves even at the rather low temperatures of 
the cluster experiments. In Fig.\ 4 (bottom frame on the right), we compare 
the $f_{\text{sh}}$'s for K$_9^+$ and K$_3^+$ calculated with the realistic 
single-particle spectrum of our FT-SCM. Note that, as a function of $T$, the 
realistic spectrum suppresses the shell effect of the heavier-than-the-trimer 
magic fragments at a faster rate than the estimate above.  

The controlling influence of the electronic entropy on the stabilization of 
electronic shells, discussed previously \cite{yl1,yl2} in studies of 
ground-state properties of clusters, has been shown here to govern the fission
patterns and their temperature dependence at the low-fissility regime
\cite{note55}. This provides the impetus for further temperature-controlled 
studies of charge instabilities in clusters.

This research is supported by the US D.O.E. (Grant No. FG05-86ER-45234).


\begin{references}
\bibitem{note2}
Previous experimentally determined $Q$-values, obtained via Born-Haber
cycles and employing a low-temperature value of the dimer dissociation
energy, are in agreement with the theoretical zero-temperature SCM results.
\end{references}
\end{document}